\documentclass[12pt,preprint]{aastex}
\usepackage{multirow}
\usepackage{hhline}
\usepackage{url}
\usepackage{epsfig}
\usepackage{amssymb}

\def\ltsima{$\; \buildrel < \over \sim \;$}
\def\gtsima{$\; \buildrel > \over \sim \;$}
\def\lsim{\lower.5ex\hbox{\ltsima}}
\def\gsim{\lower.5ex\hbox{\gtsima}}
\def\lapp{\ifmmode\stackrel{<}{_{\sim}}\else$\stackrel{<}{_{\sim}}$\fi}
\def\gapp{\ifmmode\stackrel{>}{_{\sim}}\else$\stackrel{<}{_{\sim}}$\fi}

\newdimen\minuswidth    

\setbox0=\hbox{$-$}

\minuswidth=\wd0

\catcode`@=\active

\def@{\kern\minuswidth}

\setbox0=\hbox{\rm0}
\shorttitle{}

\shortauthors{}

\begin{document} 

\title{Radio Timing and Optical Photometry of the Black Widow System PSR J1953+1846A in the Globular Cluster M71\footnote {Based on observations collected with the NASA/ESA HST (Prop. 12932), obtained at the Space Telescope Science Institute, which is operated by AURA, Inc., under NASA contract NAS5-26555.}}

\author{
M. Cadelano\altaffilmark{1,2}, C. Pallanca\altaffilmark{1}, F. R. Ferraro\altaffilmark{1}, I. Stairs\altaffilmark{3}, S. M. Ransom\altaffilmark{4}, E. Dalessandro\altaffilmark{1}, B. Lanzoni\altaffilmark{1}, J. W. T. Hessels\altaffilmark{5,6}, P. C. C. Freire\altaffilmark{7}}
\affil{\altaffilmark{1} Dipartimento di Fisica e Astronomia, Universit\`a di Bologna, Viale Berti Pichat 6/2, I-40127 Bologna, Italy }
\affil{\altaffilmark{2} INAF - Osservatorio Astronomico di Bologna, via Ranzani 1, I-40127 Bologna, Italy }
\affil{\altaffilmark{3} Department of Physics and Astronomy, University of British Columbia, 6224 Agricultural Road Vancouver, BC V6T1Z1, Canada}
\affil{\altaffilmark{4} National Radio Astronomy Observatory (NRAO), 520 Edgemont Road, Charlottesville,
Virginia 22901, USA.}
\affil{\altaffilmark{5}  ASTRON, the Netherlands Institute for Radio Astronomy, Postbus 2, 7990 AA, Dwingeloo, The Netherlands}
\affil{\altaffilmark{6}  Anton Pannekoek Institute for Astronomy, University of Amsterdam, Science Park 904, 1098 XH Amsterdam, The Netherlands}
\affil{\altaffilmark{7} Max-Planck-Institute f$\ddot{u}$r Radioastronomie, D-53121 Bonn, Germany}

\begin{abstract}
We report on the determination of the astrometric, spin, and orbital
parameters for PSR J1953+1846A, a ``black widow'' binary millisecond
pulsar in the globular cluster M71. By using the accurate position and orbital
parameters obtained from radio timing, we identified the optical
companion in ACS/Hubble Space Telescope images. It turns out to be a faint ($\rm m_{F606W}\gtrsim24$, $\rm m_{F814W}\gtrsim 23$) and variable star 
located at only $\sim0.06\arcsec$ from the pulsar timing position. The
light curve shows a maximum at the pulsar inferior conjunction and a
minimum at the pulsar superior conjunction, thus confirming the
association with the system. The shape of the optical modulation suggests
that the companion star is heated, likely by the pulsar wind. The
comparison with the X-ray light curve possibly suggests the presence of an
intra-binary shock due to the interaction between the pulsar wind and the
material released by the companion. This is the second identification
(after COM-M5C) of an optical companion to a black widow pulsar in a
globular cluster. Interestingly, the two companions show a similar light curve
 and share the same position in the color magnitude diagram.
\end{abstract} 


\section{INTRODUCTION}\label{intro}

Millisecond pulsars (MSPs) are rapidly spinning neutron stars (NSs),
formed in a binary system where, according to the canonical scenario
\citep{alpar82, batta91}, a slowly rotating NS is spun up through mass
accretion from an evolving companion star. The accretion phenomena are
usually observed during the phases of low mass X-ray binaries
(LMXBs), which can be considered as the progenitors of MSPs. The process
usually leads to a deep transformation of both the accreting and the donor
stars: the former is accelerated to millisecond spin periods, while the
latter can evolve into an intermediate anomalous evolutionary phase
\citep[see, e.g., MSP-A in NGC 6397;][]{ferraro01a}, before reaching the
final stage of a (possibly He) white dwarf \citep[WD; e.g. MSP-A in NGC
6752;][]{fer03_msp}.\\

Although the Galaxy is $\sim100$ times more massive than the entire
Galactic GC ecosystem, about 40\% of the known MSP population is found in
GCs. Such an over-abundance is indicative of a strongly
enhanced dynamical activity in these dense stellar systems. In fact, in
the Galactic field the most plausible channel for the formation of MSPs is
the evolution of primordial binaries, while in GCs dynamical interactions
promote the formation of a conspicuous number of exotic objects, such as
blue straggler stars, X-ray binaries, cataclysmic variables and MSPs
\citep{bailyn92, cool95, ferraro95, ferraro01b, pooley03, ransom05a}, which
can be used to trace the complex interplay between dynamics and stellar
evolution \citep[e.g.][]{goodman89, hut92, phinney92, ferraro95,
possenti03, fer03_dyna, ferraro09, ferraro12}. Thus, especially in the
very centre of these systems, we expect to find a large number of NSs
which are (or have been) affected by dynamical processes such as tidal
captures or exchange interactions \citep[see e.g.][]{ivanova08}. In this
respect, the study of optical companions to binary MSPs in GCs is of
outmost importance since it opens the possibility to evaluate the frequency
and time-scales of dynamical interactions in dense stellar systems, to
explore the impact of dynamics on MSP evolutionary paths and to
investigate stellar evolution under extreme conditions \citep[see e.g.][]{fer03_mass,sabbi03a,sabbi03b,mucciarelli13}.\\

Although the majority of binary MSPs have low-mass He WD companions,
recent PSR searches have considerably increased the number of
non-canonical binary MSPs. Among these, ``black widows'' (BWs) and
``redbacks'' (RBs) are of particular interest due to the presence of radio
eclipses of the MSP signal, caused by a significant amount of ionized
material ablated from the companion star because of the energy injected
by the pulsar  \citep[PSR; e.g.][]{ruderman89}. The eclipsing
regions are usually larger than the companion Roche Lobes, thus suggesting the
presence of a non degenerate and possibly bloated star, as confirmed by
several optical identifications \citep[e.g.][]{edmonds02, ferraro01a,
cocozza08, pallanca10, pallanca13, pallanca14a}. These systems are
characterized by small orbital eccentricities, tight orbits (orbital
periods $P_{orb}\lesssim1 \ \rm day$) and small mass functions, thus
implying the presence of a low mass companion. Indeed, RB companion stars
have usually masses of  $\sim0.1 - 0.5 M_{\odot}$, while BW companions are much
less massive ($M < 0.1 M_{\odot}$). Such a small value could be due to
vaporization from the strong MSP radiation and relativistic wind; thus
these systems may provide a possible explanation to the existence of
isolated MSPs, even if their observed number is too large if compared to
the expected timescale for the total ablation of the companion stars
\citep{eichler88}. The physical mechanism favoring the formation of a RB
instead of a BW is not well understood yet. \citet{benvenuto14} argued that RBs may evolve into BW systems, as a direct consequence of ablation processes, although this cannot apply to all RBs since some of them could evolve into canonical He WD systems. On the other hand, \citet{chen13} suggested that BWs cannot result from the evolution of RBs and that the discriminant factor leading to the formation of a RB instead of a BW is the reprocessing efficiency of the MSP emission by the companion star, likely related to geometrical factors.\\

Since the first eclipsing MSPs were preferentially found in GCs, it was
commonly believed that these systems form exclusively in crowded
environments \citep{king03}, mostly as a consequence of exchange
interaction, and the few objects discovered in the Galactic field were
born in GCs and later ejected. Indeed, at that time, the ratio between the
number of eclipsing MSPs and canonical MSPs was significantly higher in
GCs than in the Galactic field. This scenario has been recently altered by
the discovery of a large number of eclipsing MSPs in the Galactic field,
both in blind surveys \citep{burgay06, bates11, keith12}, and especially
in surveys aimed at the radio identification of Fermi sources
\citep[e.g.][]{ransom11, keith11}. Hence, these objects can also form in
the field from the undisturbed evolution of LMXB systems, with no need for
dynamical interactions with other stars. In such a scenario, the fraction
of eclipsing to canonical MSPs should be similar in the field and in GCs, irrespectively from their 
interaction rate per binary, since binaries in very
tight orbits are unlikely to be disrupted \citep{verbunt14}. As an example, the BWs  J1518$+$0204C and J1807$-$2459A are located respectively in a GC with a very low interaction rate (M5) and in the one with the largest rate \citep[NGC 6544;][]{pallanca14a, verbunt14}. Still, both the objects appear to be in a relatively undisturbed systems (no orbital eccentricity). Understanding the formation of BWs and RBs, including the possible role of
stellar interaction, strongly motivates multi-wavelength studies, both in
GCs and in the Galactic field.\\

Despite the importance of MSP optical companions, finding them in
crowded stellar systems like GCs is extremely challenging. Only nine companions have
been discovered so far in GCs. Three companions are He WDs \citep[see][]{edmonds01, fer03_msp,
sigurdsson03}, as expected from the canonical formation scenario, five are RBs companions \citep[see][]{ferraro01a, edmonds02, cocozza08, pallanca10, pallanca13} and only one is a BW companion
\citep[see][]{pallanca14a}. Here we report the radio timing ephemeris
of PSR J1953+1846A (hereafter M71A) in the GC M71 (NGC 6838) and the
identification of its companion star.\\

M71A is the only MSP known so far in M71 \citep{ransom05b}. M71 is a low
density GC \citep[$\log \rho_{0}=2.83$ in units of $\rm L_{\odot} pc^{-3}$;][]{harris96}, 
in a disk-like orbit \citep{geffert00}, located at
$\sim4$ kpc from the Earth. It is one of the most metal-rich clusters
among halo GCs \citep{harris96} and its surface brightness profile shows
an extended core \citep[$r_{c}=37.8\arcsec$;][]{harris96} and no
signatures of core collapse. M71A was discovered in a targeted survey
of all GCs visible with the 305-m Arecibo radio telescope
\citep{hessels07}. It is located at $\rm \alpha=19^{h}53^{m}46.42^{s} \ ; \
\delta=18^{\circ}47\arcmin04.84\arcsec$, at a projected distance of only $20\arcsec$ (0.53 core radii) 
from the cluster center  \citep{gra+10}, and it has a spin period of $\sim4.9$ ms and a low eccentricity orbit
of $\sim4.2$ hours. M71A is classified as a BW. In fact, because of its
very low mass function ($f=1.6\cdot10^{-5} M_{\odot}$), the companion is
expected to have a minimum mass of $\sim0.032 M_{\odot}$. Moreover, as
commonly found for BW systems, the radio signal shows eclipses for about
20\% of the orbital period (at 1400 MHz observing frequency), likely due
to stripped material from an evaporating companion. A Chandra X-ray
observation of this cluster revealed a source in a position compatible
with the PSR location and a luminosity of about $\rm 10^{31} \ ergs \
s^{-1}$ in the $0.3-8.0$ keV spectral range \citep{elsner08}. The light
curve is consistent with a non-steady source and the photon index ($\Gamma
= 1.89 \pm 0.32$) suggests magnetospheric radiation and/or an emission
from intra-binary shocks. In the context of an optical study of the M71 X-ray sources, \citet{huang10}
suggested as possible optical companion to M71A a star located at $\sim0.1\arcsec$ from the radio PSR and lying along the red side of the cluster main sequence (MS), in a region
commonly occupied by binary systems.
Nonetheless, its absolute magnitude
($M_{V}\sim8.5$) implies a mass of about $0.5 M_{\odot}$, inconsistent
with radio-derived mass function (in fact such a large mass would be
compatible only with a nearly face-on orbit, where no radio eclipses are
expected). Hence, \citet{huang10} concluded that this object is unlikely
to be the real companion, which could be still below the detection
threshold or, alternatively, that M71A could be a hierarchical triple
system.

In Section 2 the radio timing analysis  is presented, while Section
3 is devoted to the optical identification of the companion star. In
Section 4 we discuss the results and in Section 5 we summarize the current
knowledge about this BW system.

\section{RADIO TIMING}

Timing observations were carried out with the 305-m William E.\ Gordon telescope at the Arecibo Observatory in Puerto Rico, between MJDs 52420 (2002 May 26) and 53542 (2005 June 21), with the initial discovery observations on MJD 52082 (2001 June 22) incorporated into the timing solution.   The Gregorian L-band Wide receiver was used for the observations, sending dual-polarization data to the Wideband Arecibo Pulsar Processor \citep[WAPP; see][]{dsh00} autocorrelation spectrometers.  For most of the timing observations, 3 WAPPs were used, at frequencies centered near 1170, 1420 and 1520\,MHz,  although in the beginning, and occasionally thereafter, only one WAPP was used, and sometimes 4 WAPPs were used.  The WAPPs were configured to autocorrelate 3-level samples with 256 lags and accumulate these for 64\,$\mu$s, then sum polarizations and write the results to disk as 16-bit numbers.  See \citet{hessels07} for details of the observations.  

Offline, the PRESTO software\footnote{github.com/scottransom/presto} was used to partially dedisperse the data into 32 subbands, to reduce the data size while still facilitating searches for further PSRs in the cluster (no additional radio PSRs were found, despite careful searches of the majority of the acquired timing data).  The subbands were then folded modulo the best-known PSR ephemeris.  A Gaussian profile was fit to the summed profile from several observations for use as a standard profile, and the FFTFIT algorithm \citep{tay92} was used to determine pulse times of arrival (TOAs).  Time segments corresponding to eclipses and to the times when M71 transited at Arecibo (during which the telescope could not track the cluster) were not considered in the timing analysis.

\begin{figure}
  \begin{center}
\includegraphics[width=5in]{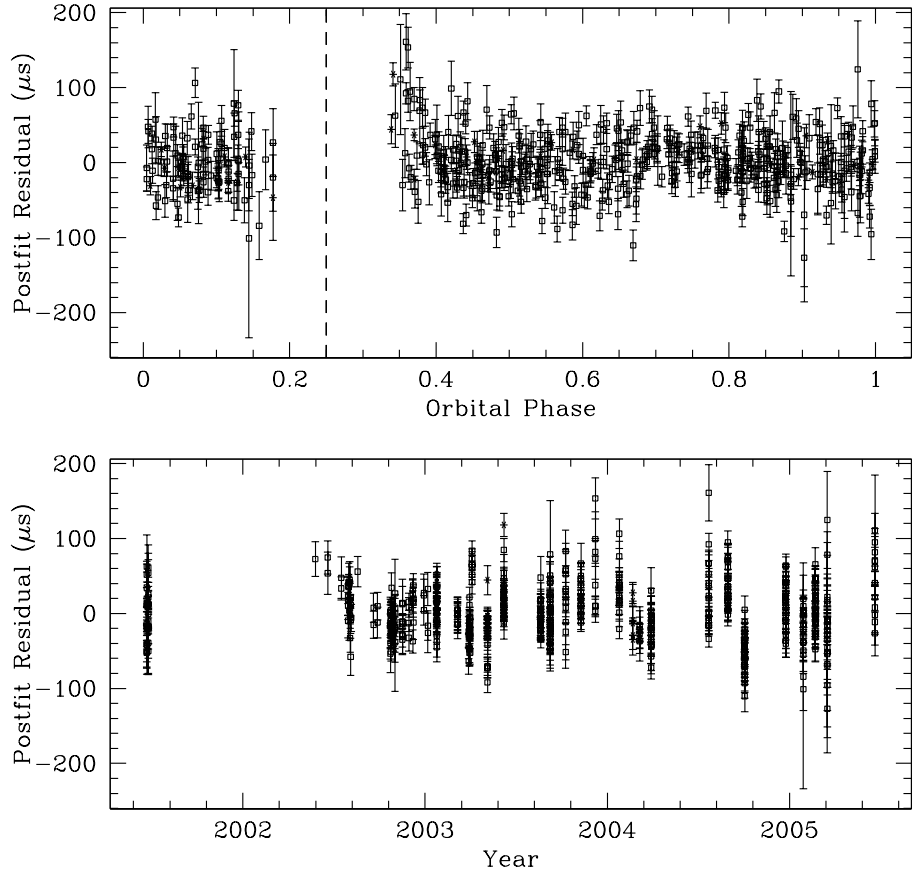}
\caption{Postfit timing residuals for M71A, as functions of orbital phase and date.  The dashed line indicates orbital phase 0.25, the PSR superior conjunction.  The asymmetric eclipse spans just over 15\% of the orbit which is typical for BW systems at radio frequencies below $\sim 2$\,GHz \citep[e.g., ][]{fst88}.}
\label{fig:M71A_resids}
\end{center}
\end{figure}

\begin{figure}
  \begin{center}
\leavevmode 
\includegraphics[width=5in]{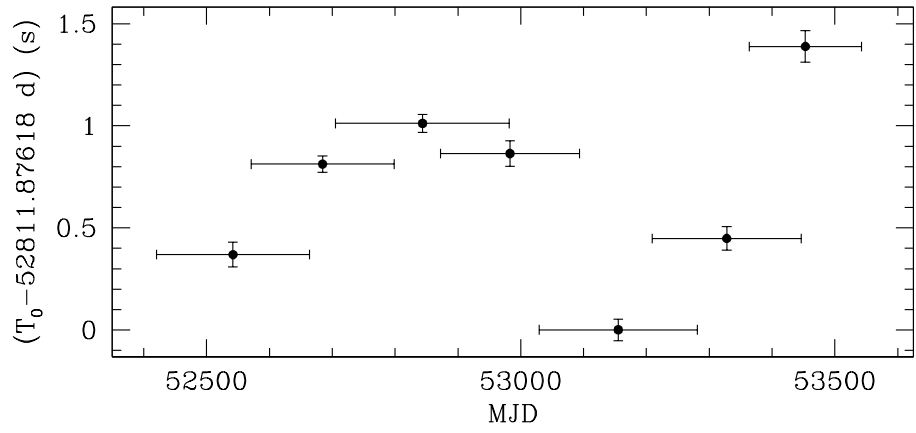}
\caption{Variation of the time of passage through ascending node (orbital phase 0), computed for overlapping segments of data and holding all other timing parameters fixed at their nominal values.}
\label{fig:t0seg}
\end{center}
\end{figure}

Timing analysis was performed with the {\sc TEMPO} software package\footnote{tempo.sourceforge.net} using the DE421 Solar System ephemeris and the TT(BIPM) clock standard.  The {\sc BT} timing model of \citet{bt76} was used, as the orbit has no significant eccentricity.  The timing parameters are listed in Table~\ref{tab:M71A_params} and residuals are presented in Figure~\ref{fig:M71A_resids}.  The root-mean-square postfit residual is 35$\,\mu$s.  The reduced-$\chi^2$ of the fit is 4.8; however, we list the parameter uncertainties as reported by {\sc TEMPO} without scaling, as the epoch-to-epoch wander in the residuals is likely due to the interactions between the two stars (see Figure~\ref{fig:t0seg}) rather than to any misestimation of the TOA uncertainties.   The high-precision radio timing position is slightly offset (0.06$^{\prime\prime}$) from the position of the optical counterpart (see Section~\ref{identification}), but agrees within the much larger uncertainty (0.2$^{\prime\prime}$) of the latter. Following the reasoning in \citet{fhn+05}, we find the maximum possible acceleration due to the gravitational field of the cluster for this line of sight to be $\pm 3.2 \times 10^{-10}$\,m\,s$^{-2}$.  This implies that most of the observed pulse period derivative ($\dot P$) is intrinsic.  Further corrections due to the differential acceleration in the Galaxy \citep[e.g.][]{nt95,rmb+14} are small.  Given the small velocity dispersion in the core of the cluster \citep[$\rm 2.3 \ km \ s^{-1}$;][]{harris96}, the velocity of the PSR relative to that of the cluster should be very small; therefore its proper motion should be very similar to the proper motion of the cluster as a whole.  A recent measurement of the proper motion of the cluster amounts to $\rm 3.93 \ mas \ yr^{-1}$ \citep{kps+13}, making the corresponding correction to $\dot P$ \citep{shk70} about half the size of that due to the Galactic acceleration.  The timing data do not allow us to derive a reliable proper motion for the PSR.  We use the range of allowed accelerations to constrain the intrinsic $\dot P$ as well as the characteristic age and surface magnetic field in Table~\ref{tab:M71A_params}.

\begin{figure}
  \begin{center}
\leavevmode 
\includegraphics*[width=5in,angle=270,origin=c]{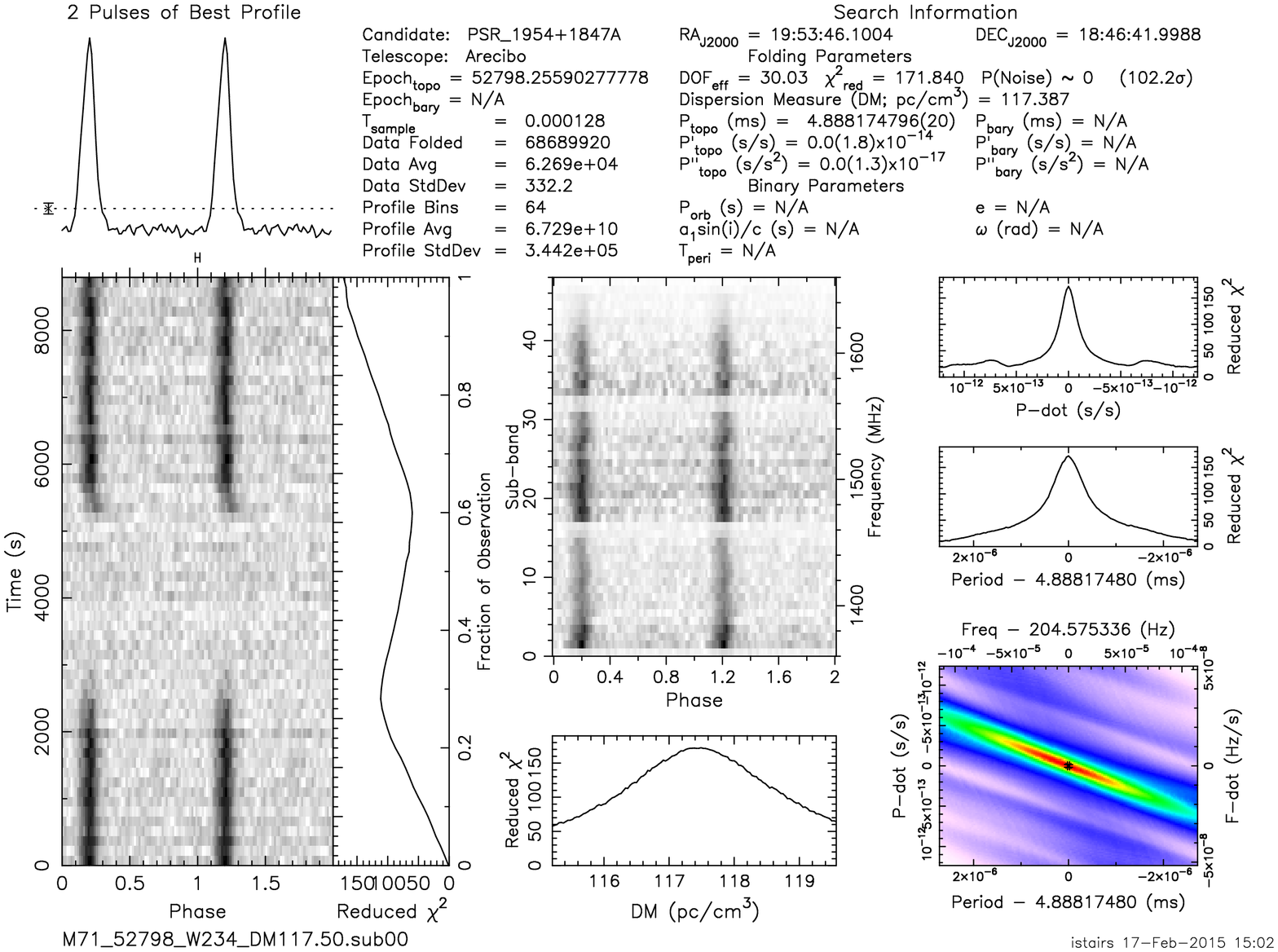}
\caption{An observation of M71A on MJD 52798 (2003 June 8), using 3 WAPPs to cover 300\,MHz of contiguous bandwidth.  The abrupt disappearance of the PSR at the start of eclipse, as well as the slight dispersive delay on reappearance, are clearly visible.  The cumulative pulse profile is plotted twice at the top of the figure. }
\label{fig:eclipse}
\end{center}
\end{figure}

The PSR is asymmetrically eclipsed between approximate orbital phases of 0.18 and 0.35, where orbital phase 0.25 represents superior conjunction.  The eclipses begin fairly abruptly but when the signal returns, it at first suffers excess dispersive delay due to ionized material within the orbit (Figure~\ref{fig:eclipse}). A discussion of the companion size and inclination angle are presented in Section 4. The mass loss from the companion star has a further manifestation in the variation of orbital parameters: Figure~\ref{fig:t0seg} shows the value of the time of ascending node passage for overlapping subsets of the data.  The variation is comparable to that seen in other black widow eclipsing systems \citep[e.g.,][]{aft94,nbb+14} and significantly less than what is typically present in the redback systems \citep[e.g.,][]{akh+13}, which have much more massive, likely non-degenerate companion stars.

\section{OPTICAL PHOTOMETRY OF THE COMPANION STAR}\label{identification}
\subsection{Observations and data analysis}\label{Sec:dataan}

The identification of the companion to M71A has been performed through two datasets of high resolution
images acquired with the Wide Field Camera (WFC) of the Advanced Camera for Surveys (ACS) mounted on
the Hubble Space Telescope (HST). The primary dataset has been
obtained on 2013 August 20 (GO12932, P.I.: Ferraro) and consists of a set of ten images in the F606W
filter (with exposure times: $\rm 2\times459 \ s ;\ 3\times466 \ s ; \ 5\times500 \ s$) and nine
images in the F814W filter ($\rm 5\times337 \ s; \ 3\times357 \ s; \ 1\times440 \ s$). We also
analyzed an archival dataset, obtained on 2006 July 1 (GO1775, P.I.: Sarajedini) with the same
instrument and same filters. It consists of four F606W images with an exposure time of 75 s and four
F814W images with an exposure time of 80 s.\\

The standard photometric analysis \citep[see][]{dalessandro08a, dalessandro08b} has been performed
on the ``flc'' images, which are corrected for flat field, bias, dark counts and charge transfer
efficiency. These images have been further corrected for ``Pixel-Area-Map''\footnote{For more
details see the ACS Data Handbook.} with standard IRAF procedures. By using the {\textrm DAOPHOT II ALLSTAR}
and {\textrm ALLFRAME} packages \citep{stetson87}, we performed an accurate photometric analysis of each
image. First of all, we modeled the Point Spread Function (PSF) by using a sample of $\sim200$ bright
but not saturated stars. The model has been chosen on the basis of a $\chi^{2}$ test and, in every
image, the best fit is provided by a Moffat function \citep{moffat69}. Then we performed a source
detection analysis, setting a $3\sigma$ detection limit, where $\sigma$ is the standard deviation of
the measured background.  Once a list of stars was obtained, we performed a PSF-fitting in each
image. In the resulting catalog we included only stars present at least in half the images for
each filter. For each star, we homogenized  the magnitudes estimated in different images and their
weighted mean and standard deviation have been finally adopted as the star mean magnitude and its
related photometric error \citep[see][]{ferraro91, ferraro92}. However, in order to perform
variability studies, for each source we also kept the homogenized magnitude measured in each frame in both
filters. Then, instrumental magnitudes have been calibrated to the VEGAMAG system
cross-correlating\footnote{We used CataXcorr, a code which is specifically developed to perform
accurate astrometric solutions. It has been developed by P. Montegriffo at INAF- Osservatorio
Astronomico di Bologna. This package is available at
http://davide2.bo.astro.it/$\sim$paolo/Main/CataPack.html, and has been successfully used in a large
number of papers by our group in the past 10 years.} our catalog with that by \citet{anderson08},
using the $\sim7600$ stars in common.\\

Since the WFC images suffer heavily from geometric distortion, we corrected the instrumental
positions (x,y) by applying the equations reported by \citet{meurer03} and using the coefficients in
\citet{hack01}. Then we transformed instrumental positions into the absolute astrometric system
($\alpha, \delta$) using the stars in common with the \citet{anderson08} catalog. The resulting
astrometric solution has an accuracy of $\sim0.14''$ in $\alpha$ and of $\sim0.13''$ in $\delta$,
corresponding to a total position accuracy of $\sim0.2''$.

\subsection{The companion to M71A}

\begin{figure*}
\begin{center}
\leavevmode
\includegraphics[width=13cm]{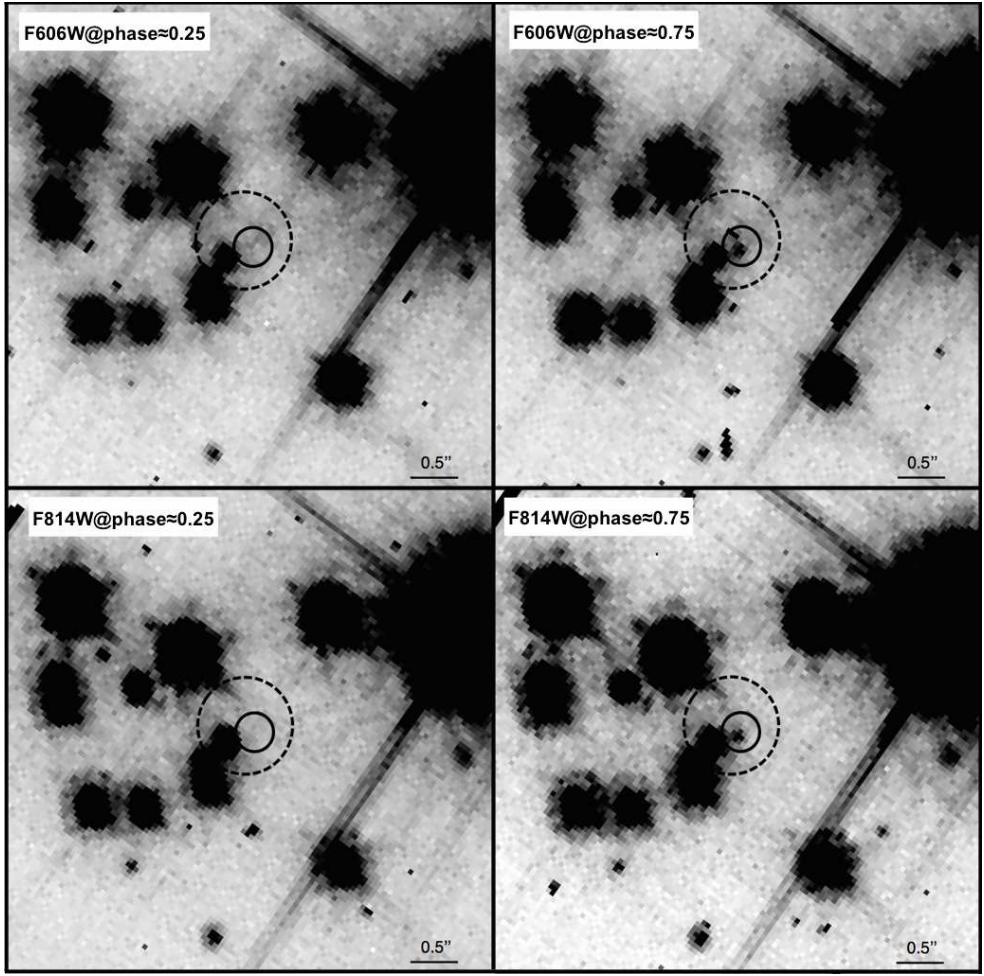}
  \caption{Primary dataset HST images of the $5\arcsec \times 5\arcsec$ region around the nominal position of M71A. The filters and the orbital phases are labeled in each panel. The solid circle is centered on the radio position and it has a radius of $0.2\arcsec$ (which is larger than the formal uncertainty from PSR timing). The dashed circle is centered on the X-ray counterpart and it has a radius of $0.5\arcsec$. The relatively bright star on the left border of the solid circle is the candidate optical companion proposed by \citet{huang10}. COM-M71A is clearly visible inside the solid circle in the right panels (corresponding to the inferior conjunction of the PSR, where the companion reaches maximum brightness), while in the left panels (at superior conjunction of the PSR) it is below the detection threshold.}
  \label{charts}
\end{center}
\end{figure*}

The search for the companion star to M71A was performed by means of an accurate photometric analysis
of all the detectable objects within a $5\arcsec \times5\arcsec$ wide region centered on the nominal
position of the MSP. Figure \ref{charts} shows the zoomed ($0.5\arcsec \times0.5\arcsec$) central
part of that region. As can be seen, a relatively bright object is found to have a position
compatible with the X-ray source (dashed circle) and the radio source (solid-line circle). This is
the star proposed by \citet{huang10} to be the optical counterpart to M71A. However, a much fainter
object, showing a strong variability, is visible in the figure. This is a quite promising object and
it is located at $\rm \alpha=19^{h}53^{m}46.4062^{s} \ ; \ \delta=18^{\circ}47\arcmin04.793\arcsec$,
only $0.06\arcsec$ from the radio position and $0.13\arcsec$ from the X-ray source, thus in
perfect positional coincidence within our positional uncertainty ($\sim0.2\arcsec$). In
the primary dataset, it has been detected in 9 (out of 10) images in the F606W filter, with a
magnitude variation ranging from $\rm m_{F606W}\approx24.3$ to $\rm m_{F606W}\approx27$, while in
the F814W filter has been detected in 6 images (out of 9) and the magnitude varies from $\rm
m_{F814W}\approx23.4$ to $\rm m_{F814W}\approx24.9$. Unfortunately, the images in the archival
dataset are too shallow to properly detect this faint object: in fact it turned out to be above the
detection threshold in only one exposure in the F814W filter.  For the four deep exposures of the
primary dataset in which the star is not visible, we estimated an upper magnitude limit by
simulating an artificial star of decreasing magnitude at the position of the candidate companion.
The derived detection threshold turned out to be $\rm m_{F606W}\sim26.5$ and  $\rm
m_{F814W}\sim25.9$.\\

\begin{figure*}
\begin{center}
\includegraphics[width=15cm]{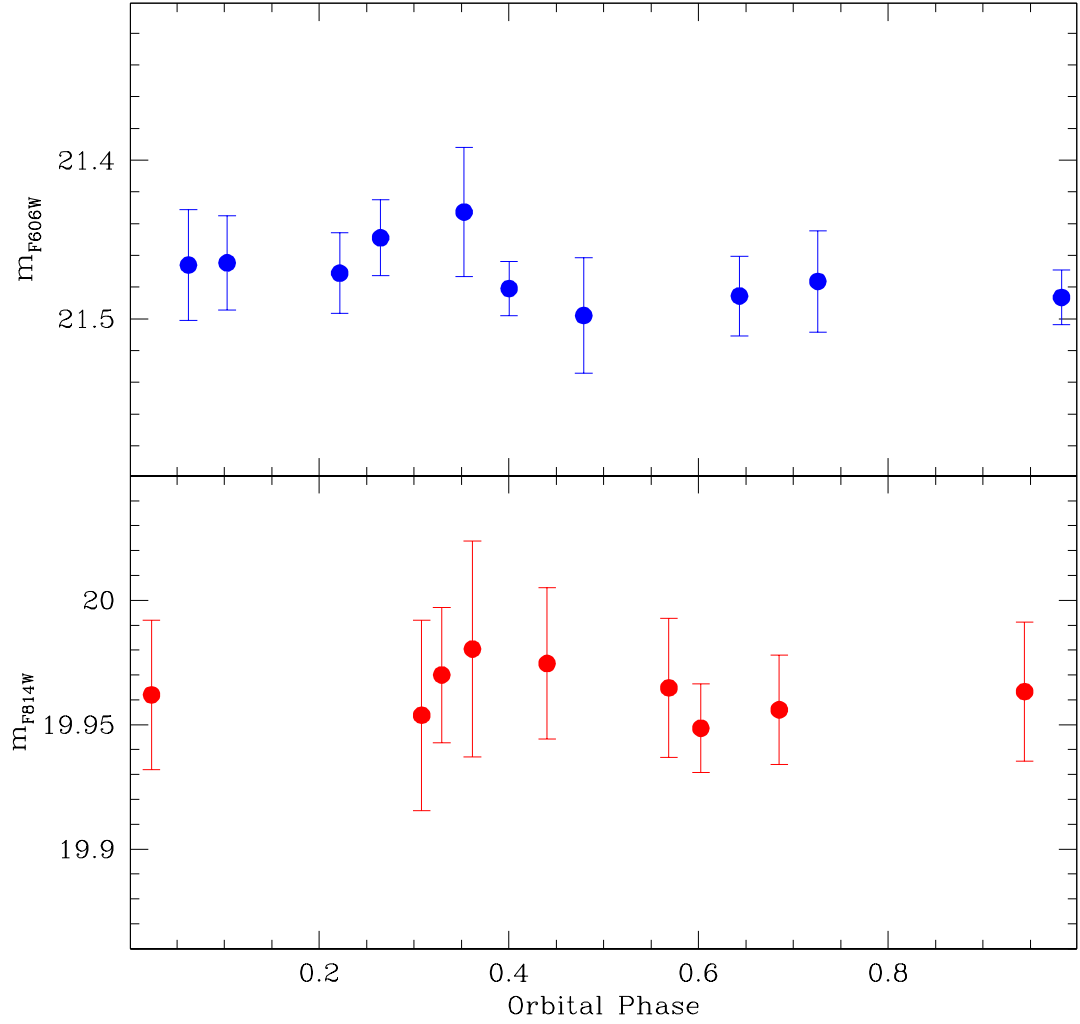}
  \caption{Light curves of the candidate companion proposed by \citet{huang10}, folded with radio orbital parameters. The absence of any magnitude modulation as a function of the orbital phase is the definitive confirmation that this object is not connected to M71A.}
  \label{curvano}
\end{center}
\end{figure*}

In order to reliably establish that the detected star is the binary companion to M71A, we
built the light curve in both the available filters by folding the optical measurements with the
orbital period and the ascending node time of the PSR (see Table~\ref{tab:M71A_params}). The results are shown in Figure \ref{curve}, and in Table 2 we report the MJD of the images with their related orbital phases and magnitudes. As
can be seen, the light curves show a sinusoidal modulation spanning at least three magnitudes and it
is fully consistent with the orbital period of the binary system. This establishes the physical
connection between the variable star and the MSP. Indeed the exposures in which the star is not
detected nicely correspond to the light curve minima. The curves have a maximum at $\phi
\approx0.75$, corresponding to the PSR inferior conjunction (where we observe the companion side
facing the PSR) and a minimum at $\phi \approx0.25$, corresponding to the PSR superior conjunction
(where we observe the back side of the companion). This behavior is indicative of a strong heating
of the companion side exposed to the PSR emission and it is in good agreement with the observed
optical properties of other similar objects \citep[e.g.][]{stappers01, reynolds07,
pallanca12, breton13, pallanca14a, li14}. For the sake of comparison, in Figure
\ref{curvano} we plot the light curve (folded following the same procedure described above) of the
possible companion suggested by \citet{huang10}. As can be seen the star does not show any
significant flux variation. \\

All these pieces of evidence suggest that the faint variable (which we name COM-M71A) is the optical
companion to the BW M71A. It is the tenth MSP optical companion and the second to a BW system in a
GC. In the color-magnitude diagram (CMD), COM-M71A is located at faint magnitudes in
a region between the MS and the WD cooling sequences, where no normal GC stars are expected. This
position is indicative of a non degenerate or semi-degenerate, low-mass and swollen star.
Interestingly, the position of this object in the CMD is quite similar to that of COM-M5C, the only
companion to a BW system known in GCs up to now and recently identified by \citet{pallanca14a} in
the GC M5.\\

\section{DISCUSSION}

\begin{figure*}
\begin{center}
\leavevmode
\includegraphics[width=15cm]{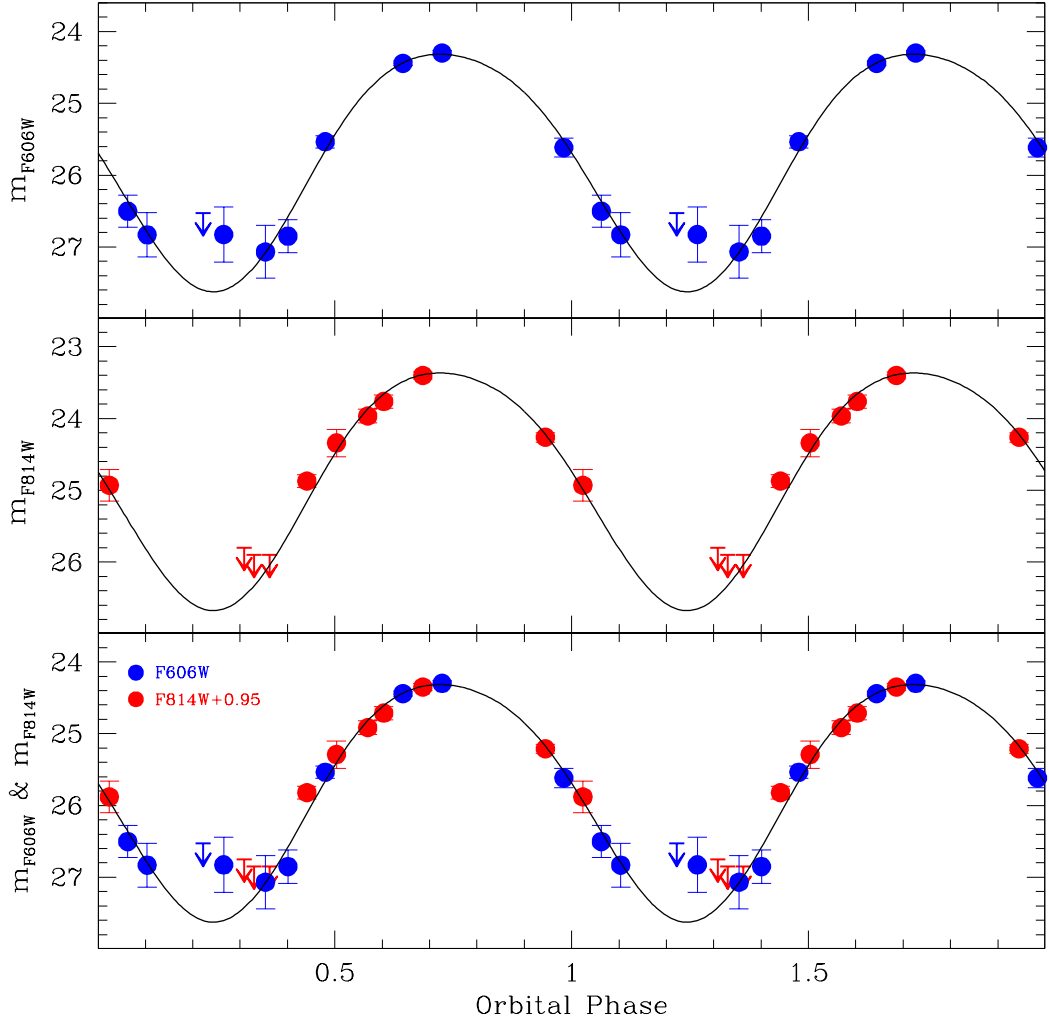}
  \caption{Light curves of COM-M71A in the F606W and F814W filters separately (upper and middle panels) and for the combination of the two (bottom panel), obtained after a 0.95 mag shift of the F814W magnitudes. All curves are folded with the radio parameters and two periods are shown for clarity. Circles mark the observed points, arrows are the magnitude upper-limits for the images where the star is below the detection threshold. The black curve in each panel is the best analytical model obtained from the combined light curve and then adapted to each filter.}
  \label{curve}
\end{center}
\end{figure*}

Since the available data do not uniformly sample the orbital phases of
the system in either the F606W or the F814W filters (see
upper and middle panels of Figure~\ref{curve}), in order to accurately determine
the light curve of the companion star we combined the two datasets
together, by applying a 0.95 mag shift to the F814W magnitudes (bottom
panel of Figure~\ref{curve}). We then used the software GRATIS\footnote{``Graphical Analyzer for TIme Series'' is a software aimed at studying stellar variability phenomena. Developed by Paolo Montegriffo at INAF-Osservatorio Astronomico di Bologna.} and a $\chi^{2}$ criterion to determine the two harmonics best fit
model\footnote{Note that a single harmonic model (i.e. a sinusoidal function) does not provide a good match of the observed light curve.}  to the curve. This is shown as a solid line in
the bottom panel of Figure~\ref{curve}.  Finally, we verified that the same
solution also provides a good fit to the light curves in each filter
separately. Indeed, the reduced $\chi^{2}$ turned out to be 1.50 for the
F606W filter, and 1.75 in F814W (see the solid curves in the upper and
middle panels of Figure~\ref{curve}).  This demonstrates that no significant
modulation of the stellar color (temperature) along the orbit is
measurable from the available dataset, and a much finer sampling of
the light curve is needed to provide additional clues on this
possibility. In Table~3 we report the maximum and minimum values
for both the magnitude and the flux in each filter, evaluated from the
best-fit model by following the procedure described in \citet{bolin12}
for the ACS. The uncertainties are calculated by using the mean
photometric errors for stars with similar magnitudes.  The magnitude
shift needed to superimpose the F814W light curve to that in the F606W
filter implies a color index of $0.95\pm0.12$ for the companion
star. By adopting a $0.54M_\odot$ WD cooling sequence from the BaSTI catalog\footnote{http://basti.oa-teramo.inaf.it} \citep{manzato08, salaris10}, this value can be converted into a
temperature of $5100\pm800$ K, which is in good agreement with those
evaluated for other BW systems \citep[e.g.][]{stappers01, pallanca12, breton13, pallanca14a, li14}.\\

\begin{figure*}
\begin{center}
\includegraphics[width=13cm]{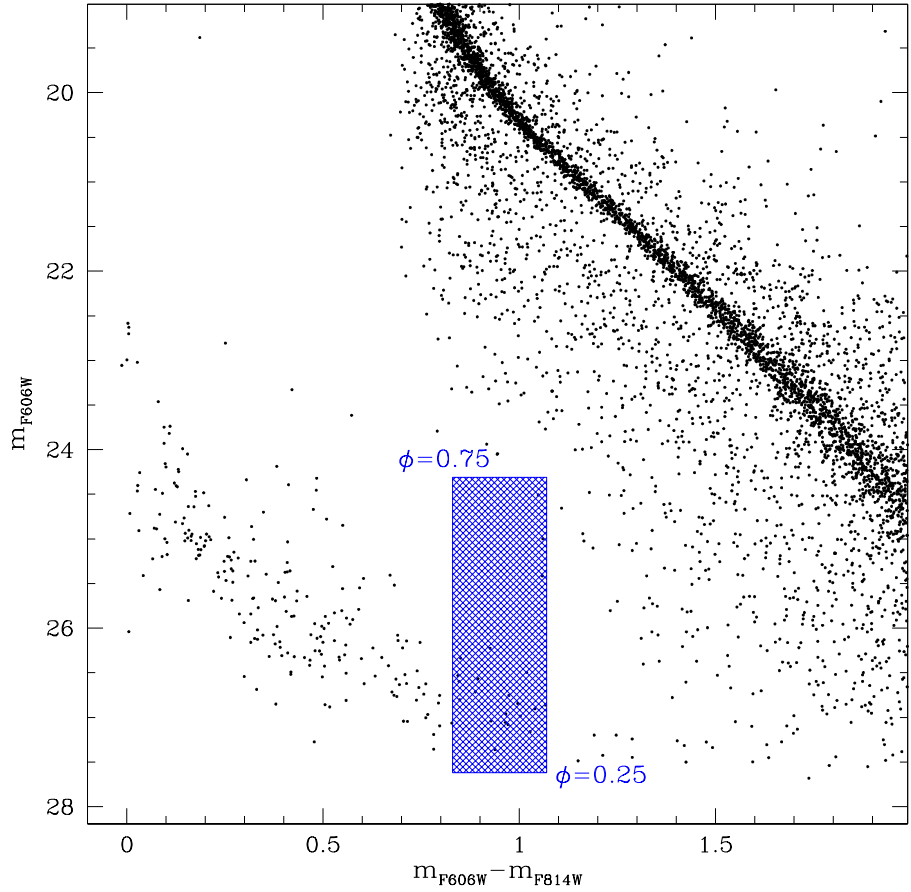}
  \caption{CMD of M71 with highlighted in blue the region occupied by COM-M71A during the whole orbital period, as predicted by the light curve model (see text, Figure~\ref{curve} and Table~3).}
  \label{cmdmotion}
\end{center}
\end{figure*}

In Figure~\ref{cmdmotion} we show the CMD, with the shaded rectangle
marking the region occupied by COM-M71A during the orbital period. The height of the rectangle
corresponds to the maximum $\rm m_{F606W}$ magnitude difference expected from the best-fit model shown 
in Figure~\ref{curve}, while the width corresponds to the photometric error at the minimum 
luminosity. As already mentioned, the star is
located between the MS and the WD cooling sequence, and it spans a range
of about three magnitudes. Of particular interest is the predicted
star position during the PSR superior conjunction, where we expect to
see the stellar side not exposed to the PSR flux (clearly this is
exactly the case only for $i = 90^{\circ}$). The CMD position of COM-M71A
in this phase could be compatible with the He WD cooling sequence,
suggesting a semi-degenerate stellar structure. However, at these low
luminosities, our analytical model is not appropriately constrained by
data. Therefore, in order to confirm this possibility, further
observations are needed. In principle, the companion mass can be constrained from the
comparison of its CMD position and theoretical isochrones.  However,
in the case of strongly perturbed stars the mass inferred in this way
can be overestimated, or suggestive of inclination angles too small to
be consistent with the presence of radio eclipses \citep[see][]{fer03_msp, pallanca10, mucciarelli13}. In our
case, not only the companion position in the CMD is clearly out of the
canonical evolutionary sequences, but also its minimum luminosity is
not properly constrained by the observations.\\

Assuming that the companion optical emission is mostly due to black-body radiation, the luminosity and temperature of this star would be consistent with an object of radius $R_{BB}\leq0.02R_{\odot}$. However, the companions to BWs usually suffer from strong tidal distortion due to the interaction with the PSR, therefore they are swollen up and possibly they can even fill their Roche Lobes. Furthermore, the presence of radio eclipses suggests that the simple $R_{BB}$ is a gross underestimate of the true stellar radius. Indeed, the Roche Lobe radius is far more appropriate to describe the size of the companion \citep[e.g.][]{stappers96, pallanca12, breton13, pallanca14a}. According to \citet{eggleton83}, the Roche Lobe radius can be computed as:

\begin{equation}
\label{RL}
\frac{R_{RL}}{a}\simeq \frac{0.49q^{\frac{2}{3}}}{0.6q^{\frac{2}{3}}+\ln\left(1+q^{\frac{1}{3}}\right)},
\end{equation}
where $q$ is the ratio between the companion and the PSR masses and $a$ is the orbital separation. Combining this relation with the PSR mass function, assuming a NS mass ranging from $1.2M_{\odot}$ to $2.5M_{\odot}$ \citep{ozel12} and an inclination angle ranging from $0^{\circ}$ to $90^{\circ}$, we find $0.22R_{\odot}<R_{RL}<1.24R_{\odot}$.\\

Interestingly, the light curve shape presents a hint of asymmetric structure in both filters: the increase to the maximum seems to be smoother than the decrease to the minimum. Despite the low statistic, this behavior could be due to a slight asynchronous rotating companion, as in the case of PSR J2051$-$0827 \citep[see][]{stappers01}. This could be the result of a tidal torque from the wind of a magnetically active star, which can result in a companion angular velocity that differs from the orbital angular velocity. Moreover, in this case the angular velocity could be subject to a variation with time due to a secular time dependence of the orbital period, due itself to a variation of the companion quadrupole moment \citep[see e.g.][]{applegate94, doroshenko01, handbook}. However, in order to probe this intriguing possibility, an uniform sampling of the light curve from new observations is needed.\\

\subsection{Reprocessing efficiency and Roche Lobe filling factor}

\begin{figure*}
\begin{center}
\includegraphics[width=15cm]{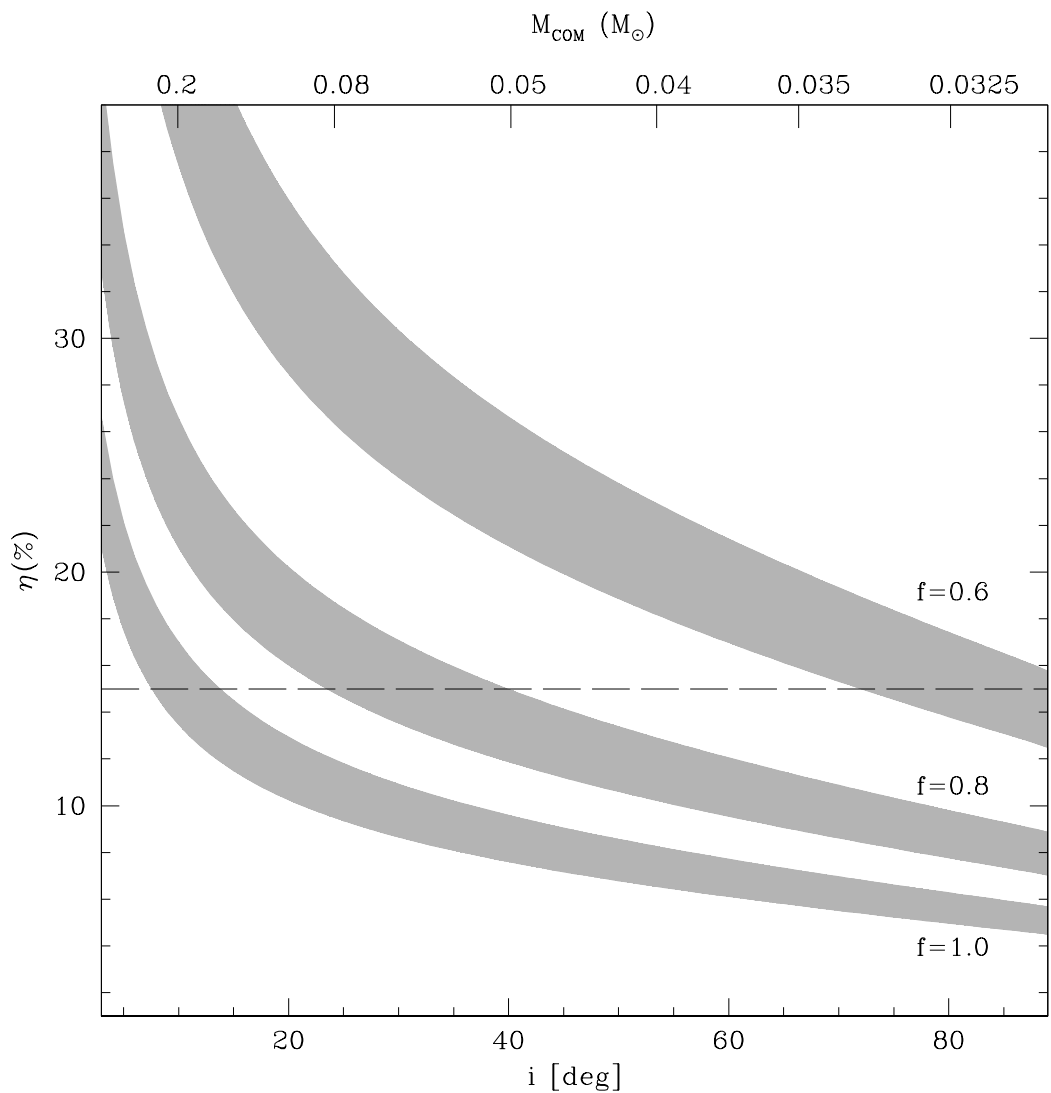}
  \caption{Reprocessing efficiency of the PSR emitted energy as a function of the inclination angle, assuming three different values of the Roche Lobe filling factor and a PSR mass of $1.4M_{\odot}$. The thickness of each strip corresponds to the range of spin-down energies measured for this PSR (see text). The horizontal dashed line at $\eta=15\%$ is a typical reprocessing efficiency reported in \citet{breton13}. On the top axis, the companion masses for a PSR mass of $1.4M_{\odot}$ are reported.}
 \label{eff}
\end{center}
\end{figure*}

Under the assumption that the optical magnitude modulation is mainly due to the heating of the
companion surface by the PSR flux, we can compare the observed flux amplitude of the light curve with the expected one ($\rm \Delta F_{exp}$) as a function of the
inclination angle $i$, by the following relation \citep{pallanca14a}:
\begin{equation}
\label{delta}
\rm \Delta F_{exp}(\it i)=\eta \frac{\dot{E}}{\it a^{2}}R_{COM}^{2}(\it i)\frac{\rm \epsilon(\it i)}{\rm 4 \pi  \it d_{PSR}^{2}}, 
\end{equation}
where $\eta$ is the reprocessing efficiency under the assumption of a PSR isotropic emission, $R_{COM}(i)=fR_{RL}(i)$ is the companion star radius, where {\it f} is the volume-averaged Roche Lobe filling factor, $d_{PSR}$ is the MSP distance, assumed to be equal to the GC distance \citep[$d_{PSR}=4.0$ kpc;][]{harris96} and $\epsilon(i)$ parametrizes the difference of the heated surface visible to the observer between maximum and minimum, as a function of the inclination angle. $\dot{E}=4\pi I \frac{\dot{P}_{int}}{P^{3}}$ is the PSR spin-down luminosity where $I$ is the momentum of inertia. Using the the spin period and its intrinsic first derivative obtained from radio timing (Table 1) and assuming $I=10^{45}\rm \ g \ cm^{2}$, we found that $\rm 4.6\cdot10^{33} \ ergs \ s^{-1}<\it \dot{E}<\rm5.8\cdot10^{33} \ ergs \ s^{-1}$, typical values within the Galactic eclipsing MSP population. Setting $\rm \Delta F_{exp}= \Delta F_{obs}$ in the F606W filter (see Table 3), we evaluated the reprocessing efficiency as a function of the inclination angle for different values of the Roche Lobe filling factor. Results are shown in Figure~\ref{eff}. As can be seen, for high inclination angles and a Roche Lobe filling companion, the reprocessing efficiency is $\sim5\%$, while for filling factor $f=0.8$ is $\sim8\%$. A typical value of 15\% \citep{breton13} would be consistent with $f\sim0.6$. Values of $f<1$ would be plausible since some works showed that BW companions not always completely fill  their Roche Lobe  \citep[e.g.][]{callanan95, stappers99, breton13}. Similar results hold for the F814W filter.\\

It is worth noting that by using $R_{BB}$ instead of $R_{RL}$ for the stellar radius, the efficiency
increases over 100\% for almost every meaningful configuration. This can be admitted only if an
anisotropic PSR emission is assumed. However, the presence of long radio eclipses and the
behavior of similar objects is a strong indication that $R_{BB}$ heavily underestimates the stellar
true radius.

\subsection{A comparison between M71A and M5C}

\begin{figure*}
\begin{center}
\includegraphics[width=15cm]{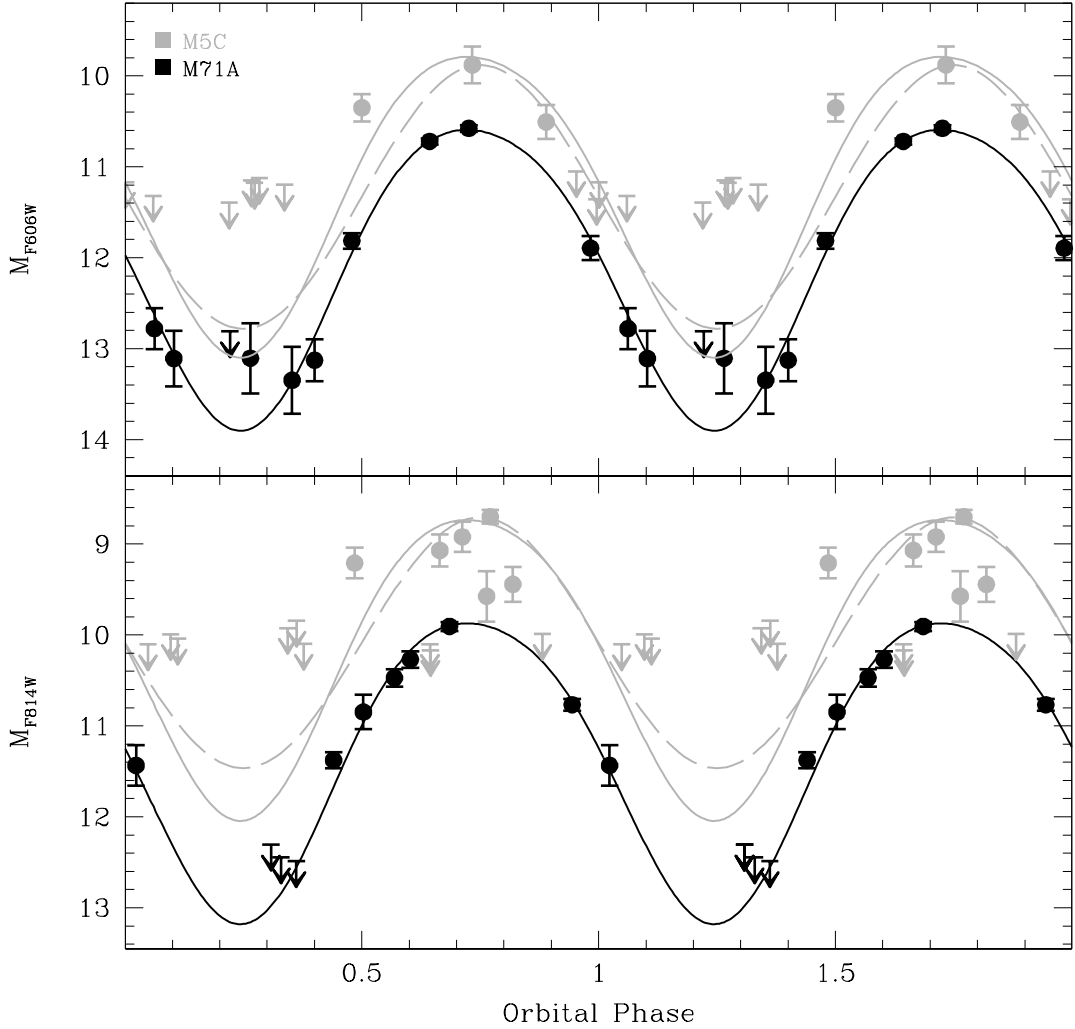}
  \caption{Optical light curves of COM-M71A (this work; black points and lines) and COM-M5C \citep[from][gray points and dashed lines]{pallanca14a}, with magnitudes reported to absolute values. The gray solid lines are COM-M71A model adapted to COM-M5C to reproduce the observed points.}
  \label{confrontocurva}
\end{center}
\end{figure*}

\begin{figure*}
\begin{center}
\includegraphics[width=12cm]{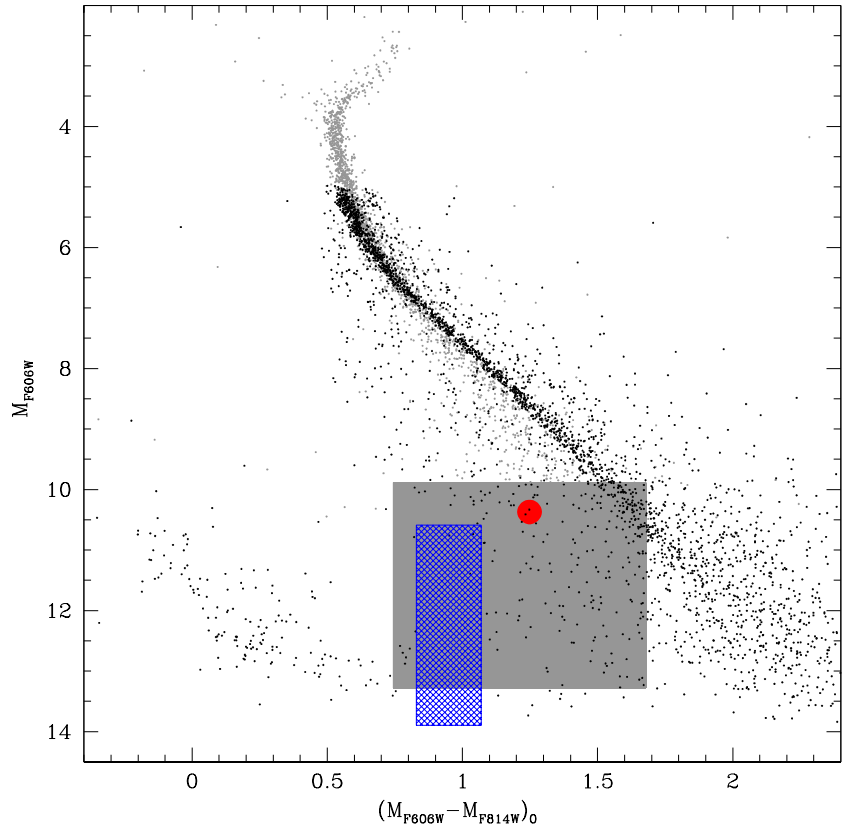}
  \caption{CMD of M71 (black dots) and M5 (gray dots). The blue shaded region is the position of COM-M71A along the whole orbital phase, as derived from the light curve model (see Figure \ref{cmdmotion}). The red point and the gray area are the indicative position of the COM-M5C \citep[see][for more details]{pallanca14a}. Both the objects are located between the MS and the WD cooling sequence, suggesting common properties of these two BW companions.}
  \label{confrontocmd}
\end{center}
\end{figure*}

So far, the optical companion to PSR J1518+0204C (hereafter M5C) was the only BW companion known in
a GC. M5C is a 2.48 ms PSR with an orbital period of $\sim2.1$ hr located in the GC M5. Its radio timing and the optical photometry of the companion star (COM-M5C) is discussed in \citet{pallanca14a}. In section 3, we anticipated some
interesting analogies between this star and COM-M71A. In order to further investigate similarities
between these BW companions, we compared the optical properties of the two objects. Figure
\ref{confrontocurva} shows their light curves, with the magnitudes reported to the absolute values.
Indeed, despite the low sampling of the COM-M5C light curves, these two objects seem to have a quite
similar optical behavior. As reference, we used COM-M71A analytical models (solid lines) for the
COM-M5C, from which we inferred that a similar light curve structure could hold even for COM-M5C,
being more appropriate than the simple sinusoid (dashed lines) used by \citet{pallanca14a}, given
the sparse number of measurements that prevent them to build an accurate model.  Figure
\ref{confrontocmd} shows the position of the two objects in the CMD. Again, considering the
uncertainties in COM-M5C magnitudes and colors, we found that they are located in the same region,
suggesting a common evolutive path for these low-mass, possibly non degenerate, swollen and heated
companions. Interestingly, in the CMD the two BW companions are located in a region completely different from that usually occupied by RBs \citep{pallanca14b}. Of course, additional identifications of BW companions are needed to firmly characterize the evolution
of these objects. In addition, using equation (\ref{delta}) for COM-M5C, setting the filling factor $f=1$ and using the spin-down period from \citet{pallanca14a} to evaluate the spin-down luminosity ($0.7\cdot10^{34} \ \rm ergs\ s^{-1}<\it \dot{E}_{M5C}<\rm3\cdot10^{34} \ ergs\ s^{-1}$), we found a reprocessing efficiency $\eta\sim5-20\%$, a value fully in agreement with what found for COM-M71A, thus further strengthening the analogies between these two systems.

\subsection{Comparing X-ray and optical light curves}

Usually, BWs with a high energy counterpart do not show any appreciable X-ray variability related to their orbital period \citep[see, e.g.,][respectively for the BWs in the GC 47 Tucanae and in the
Galactic field]{bogdanov06, gentile14}. However, this could be an observational bias, due to the
lack of deep enough and systematic surveys of BWs in the X-rays. On the other hand, it is worth noting that several RB systems clearly show orbital X-ray modulation likely due to the presence of intra-binary shocks \citep{bogdanov06, bogdanov11, bogdanov14}. M71A is an exception, since it has
been found to show periodic X-ray variability \citep{elsner08}. Very interestingly, the
determination of COM-M71A optical light curve offers the opportunity to perform a
comparison between the two. The most intriguing feature emerging from the comparison of the light  curves (both folded with the binary system parameters) is that the phase spanned by the radio eclipses ($0.18<\phi<0.35$) does not correspond to the phase of the X-ray minimum ($0<\phi<0.2$), but it nicely lines up with the optical minimum ($\phi \approx 0.25$). Thus we found that the X-ray minimum precedes the optical PSR superior conjunction. A similar effect was already observed for the RB 47TucW, a 2.35 ms binary MSP with an orbital period of $\sim3.2$ hr and a companion mass of $\sim0.15M_{\odot}$ \citep{camilo00}, whose optical light curves indicate the presence of a strong heating \citep{edmonds02}, as usually observed for BW systems. For this object \citet{bogdanov06} argue that the X-ray variability can be attributed to the presence of an intra-binary shock that is
eclipsed by the companion star. The length of the X-ray eclipse suggests that this shock is located closer to the companion star than to the MSP. In particular this behavior could be due to a
swept-back shocked region, produced by the interaction between the PSR wind and the stream of gas issuing from the inner Lagrange point L1, elongated perpendicular to the semi-major axis of the binary \citep[see][for a detailed description]{bogdanov05}. Despite the low X-ray statistics, this
is likely to be the case also for M71A, where the intra-binary shock could be eclipsed just before
the PSR superior conjunction. Even for a companion that is under-filling its Roche Lobe, this shocked region can be created thanks to the stellar wind which can result in mass outflow through L1 \citep{bogdanov05}.\\

As discussed in \citet{bogdanov05}, the Accreting Millisecond X-ray Pulsar (AMXP) SAX J1808.4$-$3658,
during quiescent states, shows several analogies with the RB 47TucW, in terms of both the X-ray
spectrum and the optical variability. Based on the discussion above, these properties are also
similar to those observed for M71A and, very interestingly, even the companion mass is comparable in
these two cases: above $0.032 M_\odot$ for COM-M71A, and $\sim 0.05 M_\odot$ for the companion to
SAX J1808.4$-$365 \citep{campana04}.  This puts M71A in the middle of the riddle, supporting the
possibility that AMXPs could be the bridge between RB and BW systems \citep{roberts14}. Clearly,
multi-wavelength studies of these objects are urged to unveil connections between AMXPs and
eclipsing MSPs, and between BWs and RBs. Indeed, several important new connections between
AMXPs and RBs have been made in the last years, especially with the discoveries of systems transitioning from one state to the other  \citep[see][]{archibald09, papitto13, patruno14,
bassa14, stappers14}.

\section{SUMMARY AND CONCLUSION}

We presented a phase-connected radio timing solution for the BW PSR J1953+1846A,
which includes a very precise position and orbital parameters determination. Taking advantage of
this precise measure of the PSR position, we have used a set of high resolution ACS/HST images to search for the companion star in the optical bands. We identified a faint and strongly variable star (COM-M71A), showing a modulation of at least three magnitudes in both used filters  (F606W and F814W). In the CMD, COM-M71A lies in the region between the cluster MS and the WD cooling sequences, thus suggesting that it is a low-mass, non-degenerate or at least semi-degenerate star, with a temperature of about 5100~K. Unfortunately, because of its faintness, it
was detectable only in 16 out of 27 images, mostly during the PSR
inferior conjunction. The light curve shows a sinusoidal shape with a period fully
consistent with the binary MSP. The maximum, during the PSR inferior conjunction, and
the minimum, during the PSR superior conjunction suggest a strong heating of the
companion star side exposed to the PSR flux. Such a behavior is in good agreement
with that observed for similar objects in the Galactic field. By modeling the light
curve, we showed that the companion reprocessing efficiency of the PSR energy is $\sim5\%$ 
for a Roche Lobe filling companion, while a typical value of $15\%$ is reasonable by assuming a filling factor of 0.6. The
comparison between the optical and X-ray light curves suggests the possible presence of
intra-binary shocks, similarly to what observed for the RB 47TucW. A
X-ray and optical follow-up will highlight the presence of this shocks and, possibly,
will allow to characterize their properties and structure. Unfortunately, the star
is too faint to allow a spectroscopic follow-up with the available instruments.
However, an optimized photometric follow-up would provide the opportunity of better
constrain the system properties, and by using, for example, phase-resolved
observations with a narrow $\rm H\alpha$ filter we could constrain the presence of
ionized material, eventually related to the intra-binary shocks.\\ 
COM-M71A is, so far, the second BW optical companion identified in a GC after COM-M5C in M5. Interestingly, both the light curve shape and the position in the CMD are quite similar in the two systems. This suggests that probably the two objects undergo a similar evolutionary path. Even though the statistic is by far too limited to draw any meaningful conclusion, at the moment no significant difference with the BW optical companions observed in the field can be evidenced, probably suggesting that no dynamical interactions are strictly needed for forming these systems.

\section{Acknowledgement}
This research is part of the project {\it Cosmic-Lab} (http://www.cosmic-lab.eu) funded by
the European Research Council under contract ERC-2010-AdG-267675.
J.W.T.H. acknowledges funding from an NWO Vidi fellowship and ERC starting grant ``DRAGNET''
(337062).

\clearpage

\begin{deluxetable}{ll}
\tablecolumns{2}
\tablewidth{0pt}
\tablecaption{Timing
parameters for PSR~J1953+1847 (M71A)\tablenotemark{a}.\label{tab:M71A_params}}
\tablehead{\colhead{Parameter} & \colhead{Value}}
\startdata
\multicolumn{2}{c}{Measured Parameters} \\ \hline
 Right ascension, $\alpha$ (J2000)\dotfill  & 
  $19^{\rm h}\,53^{\rm m}\,46\fs41966(3) $ \\
Declination, $\delta$ (J2000)  \dotfill  & 
  $+18^\circ\,47'\,04\farcs$8472(7) \\
Spin frequency $F$, Hz  \dotfill  & 
204.57006473073(3)\\
Spin frequency derivative, $\dot F$ $(10^{-15})$ (s$^{-2}$) \dotfill & $-2.0299(3)$ \\
Spin frequency second derivative, $\ddot F$ $(10^{-25})$  (s$^{-3}$)  \dotfill & 5.4(3)\\
Epoch (MJD) \dotfill & 52812.0 \\
Dispersion measure, DM (cm$^{-3}$pc) \dotfill   & 
 117.3941(15) \\
DM derivative (cm$^{-3}\mbox{pc\,yr}^{-1}$)  \dotfill  & 
$-$0.0274(17) \\
Orbital period, $P_b$ (d) \dotfill  & 0.1767950297(2)\\ 
Projected semi-major axis, $x$ (s) \dotfill  & 0.0782246(12)\\
Epoch of Ascending Node, $T_0$ (MJD) \dotfill  & 52811.8761877(3)\\ \hline
\\
\multicolumn{2}{c}{Derived Parameters} \\ \hline
Spin period, $P$ (ms) \dotfill & 4.8883007458412(6) \\
Spin period derivative $\dot P$ $(10^{-20})$ \dotfill & 4.8506(8)\\
Spin period second derivative $\ddot P$ $(10^{-29})$ (s$^{-1}$)  \dotfill & $-1.28(7)$\\
Angular offset from cluster centre $\theta_{\perp}$ ($^\prime$) \dotfill & 0.33 \\
Intrinsic period derivative $\dot P_{\rm int}$ $(10^{-20})$\dotfill & $4.3 <\dot P_{\rm int}< 5.4$\\
Characteristic age $\tau_c$ (Gyr) \dotfill & $1.4 < \tau_c < 1.8$ \\
Surface magnetic field $B_0$ ($10^8$) (gauss) \dotfill & $4.6 < B_0 < 5.2$\\
Mass function $f$ (M$_\odot$) \dotfill & 0.0000164427(8) \\
Minimum companion mass $m_c$ (M$_\odot$)\tablenotemark{b} \dotfill & 0.032\\
\enddata
\tablenotetext{a}{Numbers in parentheses are uncertainties in the last digits
quoted.}
\tablenotetext{b}{Computed assuming an orbital inclination angle of 90$^{\circ}$ and a pulsar mass of 1.4\,M$_\odot$.}
\end{deluxetable}

\begin{table}[!h]
 \begin{center}
 \title{Table 2: Optical observations of COM-M71A}\\
    \vspace{2mm} 
   \begin{tabular}{|c|c|c|c|}
     \hline
   $\phi$ & $t$ (MJD) & $\rm m_{F606W}$  &  $\rm m_{F814W}$ \\
         \hline
   0.02 &  56524.57602385  & $-$ & $24.9 \pm 0.2$ \\
   \hline
  0.06 &  56524.58290459  &  $26.5 \pm 0.2$ & $-$  \\
      \hline
   0.10 &   56524.59012681 &  $26.8 \pm 0.3$ & $-$ \\
      \hline 
   0.22 &  56524.43431532 &  $>26.5$ & $-$   \\
      \hline 
  0.26 &    56524.44193085 & $26.8 \pm 0.4$ & $-$ \\
      \hline
  0.31 &  56524.44960589 & $-$ & $>25.9$ \\
      \hline
 0.33 & 56524.63013255 & $-$ &  $>25.9$ \\
      \hline
 0.35 & 56524.45750982 &  $27.1 \pm 0.4$  & $-$ \\
      \hline
  0.36 &  56524.63586163 & $-$ & $>25.8$ \\
      \hline
 0.40 & 56524.64270200 &  $26.8 \pm 0.2$ & $-$ \\
      \hline
  0.44 & 56524.64977385 &  $-$ & $24.87 \pm 0.09$ \\
      \hline
   0.48 &   56524.65661404 &  $25.54 \pm 0.08$ & $-$ \\ 
   \hline
  0.50 & 53867.78512088$^{a}$ &  $-$ & $24.3 \pm 0.2$ \\
      \hline
   0.57 & 56524.49569959 & $-$ & $23.97 \pm 0.09$\\
      \hline
  0.60 & 56524.50166033 &  $-$ & $23.76 \pm 0.09$ \\
      \hline
 0.64 & 56524.50885348 &  $24.44 \pm 0.03$ & $-$ \\
      \hline
  0.68 & 56524.51627848 &  $-$ & $23.40 \pm 0.05$ \\
      \hline
   0.73 &  56524.52347163 & $24.30 \pm 0.03$ & $-$\\
      \hline 
  0.94 &  56524.56203070 & $-$ &  $24.26 \pm 0.06$ \\
      \hline
  0.98 &  56524.56891144  &  $25.6 \pm 0.1$ & $-$ \\
      \hline
         \end{tabular}    
         \tablecomments{Orbital phases ($\phi$), corresponding MJD ($t$) of the observations and observed magnitudes or upper-limits in both filters.\\  $^{a}$ This is the only image of the archival dataset where the companion star is above the detection threshold.  
}
       \label{tab1}
      \end{center}
   \end{table}

\begin{table}[h]
 \begin{center}
 \title{Table 3: Optical properties of COM-M71A}\\
   \vspace{2mm}
  \begin{tabular}{|c|c|c|}
     \hline
       & F606W & F814W \\
   \hline
   $\rm m_{bright}$ & $24.31\pm0.01$ & $23.37\pm0.02$ \\
   \hline
   $\rm m_{faint}$ & $27.62\pm0.09$ & $26.7\pm0.1$ \\
   \hline
   $\rm F_{bright}$  ($10^{-17} \  \rm ergs \ cm^{-2} \ s^{-1}$) & $126\pm1$ & $128\pm2$ \\
   \hline
   $\rm F_{faint}$ ($10^{-17} \  \rm ergs \ cm^{-2} \ s^{-1}$)  & $5.9\pm0.5$ & $6.1\pm0.6$ \\
   \hline
   $\Delta$F ($10^{-17} \  \rm ergs \ cm^{-2} \ s^{-1}$) & $120\pm50$  & $120\pm60$ \\
   \hline
   $\rm L_{bright}$ ($10^{29} \rm \ ergs \ s^{-1}$) & $24.2\pm0.2$ & $24.5\pm0.4$ \\
   \hline
   $\rm L_{faint}$ ($10^{29} \rm \ ergs \ s^{-1}$) &   $1.15\pm0.09$ & $1.2\pm0.1$ \\
         \hline    
    \end{tabular}  
     \tablecomments{Maximum and minimum luminosities of COM-M71A as derived from the model light curve.}
       \label{tab2}
      \end{center}
\end{table}

\clearpage

\end{document}